\documentstyle[12pt]{article}
\begin{document}
\title{ \vskip-90pt
\begin{flushright} {\normalsize CERN-TH. 6811/93} \end{flushright}
\vskip80pt \vskip12pt Is it possible to  recover information from the
black--hole radiation?}  \author{M. Schiffer\thanks{ Supported by a
John Stewart Bell Scholarship} \\ CERN, CH-1211 Geneva 23,
Switzerland}     \date{}
\maketitle
\renewcommand{\thefootnote}{}
\footnotetext{\vskip15pt \begin{flushleft}
 CERN - TH. 6811/93\\February1993
\end{flushleft}}  \renewcommand{\thefootnote}{\arabic{footnote}}
 \maketitle
\newcommand{\be}{\begin{equation}}
\newcommand{\ee}{\end{equation}}
\newcommand{\n}{\bar{n}}
\newcommand{\m}{\bar{m}}
\newcommand{\binomial}[2]{{\displaystyle#1\choose\displaystyle #2}}
\newcommand{\ie}{{,\it i.e. ,}}
\newcommand{\book}[3]{(19#1), \, {\it #2 \/} (#3).}
\newcommand{\inbook}[5]{ (19#1), \, `#2' in {\it #3 \/} eds: #4, (#5).}
\newcommand{\prd}[3]{ (19#3), {\it Phys. Rev.} {\bf D#1}, #2.}
\newcommand{\pra}[3]{ (19#3),  {\it Phys. Rev.} {\bf A#1}, #2 .}
\newcommand{\prl}[3]{ (19#3),  {\it Phys. Rev.} Lett. {\bf #1}, #2.}
\newcommand{\pla}[3]{ (19#3),  {\it Phys. Lett.} {\bf A#1}, #2.}
\newcommand{\plb}[3]{ (19#3),  {\it Phys. Lett.} {\bf B#1}, #2.}
\newcommand{\ijtp}[3]{ (19#3),  {\it Int. J, Theor. Phys.} {\bf#1},#2.}
\newcommand{\lnc}[3]{(19#3),  {\it Lett. Nuov. Cim.} {\bf#1}, #2.}
\newcommand{\nat}[3]{ (19#3),  {\it Nature}{\bf#1},#2 .}
\newcommand{\npb}[3]{ (19#3),  {\it Nucl. Phys. } {\bf B#1},#2 .}
\newcommand{\cmp}[3]{ (19#3), {\it Commun. Math. Phys.} {\bf#1}, #2.}
\newcommand{\cqg}[3]{ (19#3), {\it Class. Quant. Grav.} {\bf#1},#2.}
\newcommand{\grg}[3]{(19#3),  {\it Gen. Rel. Grav.} {\bf#1},#2.}
\newcommand{\ijmpc}[3]{ (19#3), {\itInt. J. Mod. Phys}.{\bfC#1},#2.}
\newcommand{\rnc}[3]{(19#3), {\it Rev. Nuovo Cimento} {\bf#1}, #2}
\newcommand{\WBox}{\framebox[2mm]{\rule{2mm}{0mm}}}
\newcommand{\RBox}{\raisebox{1.1mm}{\framebox[2mm]{\rule{2mm}{0mm}}}}
\newcommand{\mi}{{\mbox{\tiny min}(m,n)}}

\date{}
\begin{abstract}
In the framework of communication theory, we
analyse the gedanken experiment in which beams of quanta bearing
information are flashed towards a black hole. We show that  stimulated
emission at the horizon provides a correlation between incoming and
outgoing radiations consisting of bosons. For fermions, the mechanism
 responsible
for the correlation is the Fermi exclusion principle. Each one of these
mechanisms is responsible for the a partial  transfer of the information
 originally
coded in the incoming beam to the black--hole radiation. We show that
 this process
is very efficient whenever stimulated emission overpowers spontaneous
 emission
(bosons). Thus, black holes are not `ultimate waste baskets of
 information'.
\end{abstract}
\renewcommand{\thepage}{}
\newpage
\addtocounter{page}{-1}
\renewcommand{\thepage}{\arabic{page}}
\section{Introduction}
In the wake of Hawking's seminal paper in which he proved that black
holes radiate with a (distorted)  black-body spectrum \cite{hawking},
a fundamental question touching the basis of quantum mechanics
emerged.

The transmission of information by means of black-body radiation is
thermodynamically forbidden. Therefore, it is widely believed that all
information stored in a physical system is inexorably lost as it
crosses the black--hole event horizon because, so it seems, it cannot
be recovered from Hawking's  radiation. Accordingly,  as a black hole
evaporates completely pure states could be  converted into mixed ones
(thermal radiation), threatening the very fundamentals of quantum
mechanics that predict unitary evolution of quantum states.

There are many approaches to the problem. Hawking \cite{hawking,dollar}
advocates that all information regarding the black--hole past history
is lost forever and that quantum mechanics has to be reformulated to
accommodate this fact. Others, believe that this information remains
stored inside a black hole until the last moments of evaporation. Then
either (i) the black hole stabilizes at some radius of the order of
Planck's length and all the information in question is retained in its
interior \cite{remnant} or (ii) all this information is
instantaneously liberated to the environment
\cite{remnant1}--\cite{remnant3}. However, any of these scenarios
requires a huge amount of information to be confined within a tiny
region of space--time, something against our intuition and in conflict
with the entropy bound formulated some years ago by Bekenstein
\cite{bound,bekschi}. Another group believes that the resolution of
the paradox lies in the physics of superstrings:  black holes are
assumed to have some quantum W-hair that, in principle, could be
detected via Bohm--Aharonov experiments \cite{nick1}--\cite{nick3}. A
more conservative stand was taken by 't Hooft \cite{hooft,hooft'} who
suggested that the information in question leaks from a black hole  by
some yet unknown mechanism that correlates the outgoing and incoming
radiations. In this direction, Bekenstein \cite{bekenstein} very
recently explored the fact that the coefficient of transmission
through the potential barrier that surrounds a black hole is not unity
(Hawking's radiation is not exactly black body), to show that  there
is enough  room from a thermodynamical point of view for the black
hole to leak all the information it stored along its past history.

Our task in this paper is by far less ambitious than the scope of the
involved paradox. Here, we shall consider the gedanken experiment where
information is coded in beams of quanta (very much as is done inside
optical fibres used for telephonic communication), which are then
flashed towards the black hole. Common wisdom asserts that as the beam
crosses the horizon all the information it bore is lost forever.
However, this neglects a fundamental aspect of black-hole radiance:
the approach of the incoming beam at the horizon is followed by the
stimulated emission  of other quanta. In the framework of
communication theory  we shall prove that, thanks to this stimulated
emission, information coded in an incoming beam consisting of bosons
 is partially
transferred to the outgoing radiation and that this process is very
efficient for all modes satisfying $\hbar \omega << T_{\mbox{\tiny
BH}}$, provided that the mean number of quanta in the incoming beam
$\n>>1$, because stimulated emission then overpowers spontaneous
emission. Under these conditions {\it most of the information
originally coded in the ingoing beam can be recovered from Hawking's
flux}. In the case of a beam consisting of fermions, the exclusion
 principle
provides the mechanism responsible for a similar correlation.

\section{The role of stimulated emission} An isolated black hole emits
(spontaneously) bosons with a spectrum  \be \n=\frac{\Gamma}{e^x - 1}
\label{spontaneous} \ee where $x=\hbar \omega/T_{\rm bh}$ and $\Gamma$
is the  coefficient of transmission through the potential barrier  that
surrounds the black hole \cite{hawking}, a function of black hole and
field-mode parameters  which cannot be cast in a closed form. For this
reason, it became a widespread practice to  set $\Gamma=1$ while
discussing black--hole radiance, mainly because it is believed that
the essence of the effect is captured  by the thermal factor of this
expression alone.

This   practice gives the misleading impression that black holes are
inert to incoming radiation. But they are not. Suppose that a black
hole is impinged by $n$ quanta. Then, the mean number of quanta in the
outgoing flux is composed of spontaneous emission [(eq.
\ref{spontaneous})] and a fraction $(1-\Gamma)$ of the original beam.
For incoming thermal radiation at temperature $T$
\cite{meisels,centenial}:      \be  \n=\frac{\Gamma}{e^x - 1} +
\frac{1-\Gamma}{e^y-1}  \ee  where $y=\hbar \omega/T$. On the other
hand, the conditional probability $p(m|n)$ that the black hole returns
$m$ quanta given that $n$ are incident  (in a given mode) is defined
via the equation  \be  p_o(m) = \sum_{n=0}^{\infty} p(m|n) p_i(n) \, ,
\ee where $p_i(n)$ and $p_o(m)$ are the probabilities that $n$ quanta
are incident and $m$ emitted, respectively. This conditional
probability can be extracted from the above two equations using
maxentropy techniques \cite{meisels}:     \begin{eqnarray}   p(m|n)
&=& \frac{(e^x-1)e^{nx}\Gamma^{m+n}}{e^x-1+\Gamma}
\sum_{k=0}^{\mbox{{\tiny min}}(m,n)}
\frac{(m+n-k)!}{k!(n-k)!(m-k)!}X^k  \nonumber \\ & &X=2
\frac{1-\Gamma}{\Gamma^2} (\cosh x -1) -1 .  \label{conditional}
\end{eqnarray} Similar results were  also produced by field theoretic
calculations \cite{wald}. Although this conditional probability is a
quite complicated distribution, Bekenstein and Meisels \cite{meisels}
disentangled it into distributions corresponding to  three different
processes, namely, elastic scattering, spontaneous and stimulated
emissions: \be p(m|n) = \sum_{k=0}^\mi p_{\mbox{\tiny scat}}(k|n)
p_{\mbox{\tiny spont+stim}}(m-k|n) .   \ee In this convolution, the
first factor stands for the probability that $n-k$ quanta are absorbed
while $k$ are scattered:  \be p_{\mbox{\tiny
scat}}(k|n)=\binomial{n}{k} \Gamma_o^{m-k} (1-\Gamma_o)^k \, , \ee
where  \be \Gamma_o=\frac{\Gamma}{(1-e^{-x})} \ee and $1-\Gamma_o$
stand respectively for the absorption and (elastic) scattering
probabilities of one quantum. The binomial factor takes care of the
correct statistics for many bosons.

The second factor  in the convolution  \be
 p_{\mbox{\tiny stim+spon}}(m|n)
=\binomial{m+n}{m}(1-e^{-\gamma})^{n+1} e^{-\gamma m}  \label{ss} \ee
represents spontaneous {\it and} stimulated emission  as shown by
evaluation of the mean number of returned quanta for a fixed number
$k$ of incoming ones:  \be \m_{\mbox{\tiny spont+stim}} =  \sum_m
p_{\mbox{\tiny spont+stim}}(m|k)  m = \frac{1}{e^\gamma-1}(k+1) \, .
\label{return} \ee Whenever stimulated emission takes place, we expect
a  black hole to behave very much like a laser, producing
amplification of the incoming beam (signal) with negligible
degradation of the information it originally bore.

In what follows, we shall analyse this question in the context of
communication theory \cite{review,yamamoto}. Assume that the actual
state $|n\rangle$ the incoming  field mode is in  is not known a
priori, only its occurrence probability $p_i(n)$. The amount of
ignorance concerning the signal's actual state is Shannon's entropy
\cite{shannon}:    \be H_i = - k \, \sum_n p_i(n) \ln p_i(n) \, ,
\label{hi} \ee The constant $k$ fixes the units of information: for
$k=1/\ln 2 $ it is measured in bits, etc. Upon detection of the signal
an observer  picks up one from all possible states, gaining an amount
of information equal to Shannon's entropy.

{}From the point of view  of communication theory, a black hole acts as a
source of noise,  jamming the information borne by the original
signal. In the presence of this noise, the outgoing radiation is
associated with a larger entropy than is the incoming radiation,
because this noise introduces a further measure of uncertainty in the
signal. Nonetheless, this larger entropy does not correspond to a
larger amount of useful information, i.e. the one that, in principle,
could be recovered at the output. Thus, after reaching the horizon
$H_i$, no longer represents the information borne by the signal
because this has since been adulterated by noise. The procedure for
dealing with  this situation was outlined by Shannon \cite{weaver} who
noted that $H_{i|o}$, the conditional entropy of the input {\it when
the output is known\/}:    \be  H_{i|o} \equiv - \sum_{n,m} p(m|n)
p_i(n)\, \ln \left[\frac{p(m|n)p_i(n)}{p_o(m)}\right]  \label{Hio}
\ee must represent the extra uncertainty introduced by the noise,
which hinders reconstruction of the initial signal when the output is
known. Thus, he interpreted     \be H_{\mbox{\tiny{useful}}}= H_i -
H_{i|o} \label{useful} \ee to be the useful information, meaning the
one which we could, in principle,   recover from the output signal,
{\em even in the face of noise}. Thus, $H_{\mbox{\tiny{useful}}}$
represents the informational content of the outgoing radiation. We can
also regard  \be H_{o|i} \equiv - \sum_{n,m} p(m|n) p_i(n)\, \ln
p(m|n)  \label{Hoi}  \ee  as the uncertainty in the output {\it for a
given input\/}, as the effect of the noise.  By means of Jannes
identity, it is trivial to show that $H_{\mbox{\tiny{useful}}}$ can be
alternatively expressed in terms of $p_o(m)$
 \be H_{\mbox{\tiny{useful}}}= H_o - H_{o|i} \, . \label{useful'} \ee

Now, we wish to code information in the incoming beam in the most
efficient way in order to optimize its transfer. In other words, we
wish to maximize $H_{\mbox{\tiny{useful}}}$ with respect to either
$p_i(n)$ or	 $p_o(m)$. The variation of this quantity for a fixed mean
number of quanta gives \cite{wash}   \be p_o(m) = e^{-(\alpha +\beta m
+ B(m) )} \, . \label{ansatz}  \ee  where the `chemical potential'
$B(m)$ depends on the conditional probability through the equation:
 \be  \sum_m B(m) p(m|n) = - \sum_m p(m|n) \ln p(m|n) \, . \label{B}
\ee  Inserting eq. (\ref{ansatz}) into eq. (\ref{useful'}), we obtain
the amount of information that can be transmitted in the presence of
noise in the optimal regime,   \be  I_{\rm max}= \alpha + \beta \m .
\label{imax}  \ee  In the above,  $\alpha$ and $\beta$  should be
determined by normalization and mean number of quanta conditions.
Namely,  \be  \alpha = \ln\sum_m e^{- ( \beta m + B(m))}
\label{alpha}  \label{normalisation} \ee  and  \be  \m =
-\frac{\partial \alpha}{\partial \beta} \, . \label{mean}  \ee Shannon
proved the important result that the optimal regime for information
transfer can always be achieved in practice by means of an efficient
coding of the message, but that it can never be exceeded: any
information we try to send in excess of $H_{\mbox{\tiny{useful}}}$
will be washed out by noise \cite{weaver}.

 We are now in a position to calculate the amount of information borne
by the outgoing radiation in the optimal regime. Before doing so, it
is  worth recalling that the geometry becomes transparent ($\Gamma
\rightarrow 1$) for very large frequency modes. In this limit,
$X\rightarrow -1$ and the sum in eq. (\ref{conditional}) is unity.
Thus, the outgoing radiation is purely thermal  \be  p(m|n)
\rightarrow  (1-e^{-x}) e^{-mx} \, . \ee Notice that since the
conditional probability is independent of $n$, $p_o(m) \rightarrow
p(m|n)$. Inspecting  eq. (\ref{Hio}) we see that in this limit
$H_{i|o} \rightarrow H_{i}$ and consequently $H_{\mbox{\tiny useful}}
\rightarrow 0$. This result could be foreseen from thermodynamical
considerations because the second law of thermodynamics forbids
thermal radiation to convey any information.

In order to render these calculations feasible, we will have to resort
to a simplification  of the problem by setting $\Gamma_0=1$, which
corresponds to omitting scattering  processes. In other words, we
assume that all incoming quanta are absorbed  by the black hole.
Accordingly, in so doing we shall be setting a lower bound on the
amount of information that could be borne by the outgoing radiation,
which we hope wil be very close to its actual value. With this
simplification, the relation between $x$ and $\gamma$ becomes very
simple: \be \gamma=\ln(e^x+1) \, . \label{gamma} \ee Thus, our task
now is to solve eq. (\ref{B}) for the distribution (\ref{ss}). This
calculation is shown in the appendix and the result is: \be B(m)=
e^{\gamma m} (-1)^m \sum_{k=m}^{\infty}
\binomial{k+1}{m+1}\sum_{n=0}^{k} F(n+1)(-1)^n \binomial{k}{n}\, ,
\label{inverse} \ee where \be F(n)\equiv - \sum_{m=0}^{\infty}
\binomial{m+n}{n} e^{-\gamma m} \ln p(m|n) \, . \ee Calculating $F(n)$
entails a summation of the logarithm of a binomial [see eq.
(\ref{ss})], a task that we were not able to accomplish analytically.
Since the logarithm of a binomial, regarded as a function of its lower
argument, is very close to a parabola [see fig. I] we adopted the
following approximation \be \ln \binomial{m+n}{n} \approx 4 \ln 2
\frac{m n}{m+n} \ee and carried out the summation detailed in the
appendix. The final result for $F(n)$ is \be F(n) \approx
(n+1)\frac{\gamma e^{-\gamma}-(1-e^{-\gamma})\ln (1-e^{-\gamma})}
{(1-e^{-\gamma})^{n+2}} -  n \,\frac{(4\ln 2)
e^{-\gamma}}{(1-e^{-\gamma})^{n+1}} \, . \ee Inserting back this
result into eq. (\ref{inverse}), we obtain for $B(m)$: \be B(m)
\approx \mu m + \nu \ee where \begin{eqnarray}
\mu&=&\gamma-(e^\gamma-1)\ln(1-e^{-\gamma}) - 4 \ln 2
(1-e^{-\gamma})\, \mbox{and}\nonumber\\ \nu&=&(4 \ln 2) e^{-\gamma} \,
. \label{munu} \end{eqnarray} Fixing $\alpha$ and $\beta$ through eqs.
(\ref{normalisation}) and (\ref{mean}) in terms of $\m$ one obtains
\be I_{\mbox{\tiny max}}(x) \approx \ln(\m+1) + \m
\ln\left(\frac{\m+1}{\m}\right) -\nu-\m \mu \, . \ee Recalling the
definitions of $\mu$ and $\nu$ [eq.(\ref {munu})]; the relation
between the mean numbers of incoming and outgoing quanta [eq. (\ref
{return})]; and the relation between $\gamma$ and $x$
[eq.(\ref{gamma})], we plotted $I_{\mbox{\tiny max}}$ against $x$ for
$\n=1,10,20,100$ (see fig. II). The horizontal line represents the
information originally coded in the incoming wave. Notice that for
$\n$ sufficiently large,  $I_{\mbox{\tiny max}}$  develops an
unphysical negative tail, which must be a consequence of our (crude)
strategy of approximating the logarithm of a binomial by a parabola.
At any rate, there are features that are  likely to be universal.
First, for high--frequency modes $x>>1, \,I_{\mbox{\tiny
max}}\rightarrow 0$, because in this limit thermal radiation
overpowers stimulated emission ($\Gamma \rightarrow 1$). The other
feature is that as the mean number of incoming quanta $\n$ grows, less
and less information is degraded for all modes $x << 1$.
 This makes the case for
the following picture: whenever stimulated emission overpowers
 thermal radiation,
the amount of information that can be recovered from the black-hole
 radiation is
close to that originally coded in the incoming one.

Now, what happens if the incoming beam contained fermions instead of
 bosons?
At  first thought we are led to say that the information carried by
 fermions
is completely washed out by the black hole because the mechanism,
responsible
 for correlating incoming and outgoing radiations, that worked for
bosons,
 cannot take place (the exclusion principle forbids stimulated emission of
fermions). Fortunately, this is not quite true.

\section{The role of the exclusion principle}

Suppose
 that a black hole  spontaneously radiating fermions with a (distorted)
thermal
 distribution is hit by a fermion, which is then scattered. In order to
conform
 to the exclusion principle,
  the black hole response must be to suppress its
own
 radiation
  of a similar fermion, leaving a definite imprint in the outgoing
radiation.
 In other words, the exclusion principle  provides the mechanism that
correlates
 fermion incoming and
  outgoing radiations. As we shall see, this mechanism
will
 be
  responsible for the a partial transfer of the information stored in the
beam to the outgoing radiation.

The
  mean
   number of fermions in the outgoing radiation is, as before, composed of
the
 spontaneous emission and of a fraction $1-\Gamma$ of the  mean number of
incoming fermions,

\be
\m=\frac{\Gamma}{e^x+1} + (1-\Gamma) \n \, .
\ee
{}From
 this equation it is possible to read the conditional probability 
\cite{meisels} (notice that here $m,n$ take only the values $0,1$):
\begin{eqnarray}
p(0|0)&=&1-\frac{\Gamma}{e^x+1}\, ; \nonumber \\
p(1|0)&=&\frac{\Gamma}{e^x+1}\, ; \nonumber \\
p(0|1)&=&\frac{\Gamma}{e^{-x}+1} \,\,\, \nonumber
\end{eqnarray}
and
\be
p(1|1)=\frac{\Gamma}{e^x+1}+(1-\Gamma)\,
\ee
In order
 to find the optimal information transmission regime we follow the same
steps we
 took in the previous sections. First we have to solve eq. (\ref{B}) with
the above distribution. For fermions there are only two $B$'s:
\be
B(0)=\frac{a\, p(1|1) - b\, p(1|0)}{1-\Gamma}
\ee
and
\be
B(1)=\frac{b\, p(0|0) - a\, p(0|1)}{1-\Gamma}\, ,
\ee
where
\be
a=-p(0|0) \ln p(0|0)-p(1|0) \ln p(1|0)
\ee
and
\be
b=-p(0|1) \ln p(0|1)-p(1|1) \ln p(1|1) \, .
\ee

The optimal rate can be expressed in terms of these $B$'s:
\be
I_{\mbox{\tiny max}}= -\left[(1-\m)\ln(1-m) +\m\ln\m\right]
-B(0) +\left(B(0)-B(1)\right)\m \, ,
\ee
The
 unique piece missing
  is the transmission coefficient. We borrowed the low--
frequency
 limit of $\Gamma$ for a massless fermion with $l=s$ from Page's work on
black-hole emission rates \cite{page}
\be
\Gamma \approx (\frac{x}{8\pi})^2 \, \, , x < 1
\ee
and plotted
 the graphic of $I_{\mbox{\tiny max}}$ for $\n=0.5$ and $0\leq x<\leq
1$ in Fig. III. Since a fermion is a binary system {\em par excellence},
information
 is here measured in bits. Observe that at $x \approx 1$, the limit
where Page's
  approximation breaks down, most of the information stored in the
incoming beam is transferred to the outgoing radiation.

\section{Concluding remarks}
In this paper we showed that, contrary to common wisdom,
black holes are not `ultimate waste baskets of  information'.
When hit by radiation consisting of bosons, the black-hole response
consists of
 stimulated emission of other quanta. For fermions the mechanism is
entirely
 different. In order to conform to the exclusion principle, the black hole
must suppress
 its own spontaneous emission of a fermion having  the same quantum
numbers
 as the one it scattered. Both mechanisms provide a correlation between
outgoing
 and ingoing modes, which allows  information originally stored in the
incoming
 radiation to be partially transferred to the  radiation emitted by the
black hole.

Unfortunately, this mechanism is not  efficient enough to
resolve the black--hole--information paradox because
 thermal radiation overpowers stimulated emission of bosons for the vast
majority
 of modes. For fermions, whenever $x>>1$, the spontaneous emission is not
affected because
 all incoming quanta are absorbed $ (1-\Gamma) \rightarrow 0$.
If we
 wish to solve the `black--hole information paradox' from a conservative
standpoint
 like 't Hooft's, we have to search for new mechanisms that could account
for a perfect correlation between incoming and outgoing radiation.

{}From a technical point of view, the crude approximation adopted for the
logarithm of the binomial has to be overcome, as well as the
negligence of scattering processes by the geometry. From a more
fundamental standpoint, there also remains
to  be investigated the transmission of information by means of
superradiant modes. Since in these  modes outgoing and incoming
radiations ought to be perfectly correlated, no information is
expected to be degraded by the black hole there.

Recently two-dimensional dilatonic black holes became objects of
intense research \cite{giddings} because they allow back reaction
effects to be taken into account. Let us recall that in two dimensions
the coefficient $\Gamma \equiv 1$. According to our informational
theoretic approach and the recent work by Bekenstein
\cite{bekenstein}, two--dimensional, in contrast to four--dimensional
black holes, do behave as perfect sinks of information. Thus, the
physics of two-- and four--dimensional black holes are entirely
different, and it might well be possible that all technology learned
in two dimensions might be useless for the real case of
four--dimensional space--time.

\vskip1cm

{\bf Acknowledgements}: I am thankful to the World Laboratory for
partial financial support and J. D. Bekenstein for enlightening
discussions.

\appendix \section*{Appendix: The inverse matrix of a binomial}

Inserting the distribution (\ref{ss}) into the equation for $B(m)$ [eq.
(\ref{B})] and cancelling an overall factor $(1-e^{-\gamma})^{n+1}$
yields \be \sum_{m=0}^{\infty} B(m) e^{-\gamma m} \binomial{m+n}{n} =
F(n)\, , \label{F} \ee where \be F(n)=-\sum_{m=0}^{\infty}
\binomial{m+n}{n} e^{-\gamma m}\ln p(m|n) \, . \ee Solving eq.
(\ref{F}) for $B(m)$ is equivalent to the problem of finding the
inverse matrix of the binomial (infinite matrix).  To this end, let us
multiply both sides of this equation by $z^{-n}$, where $|z|\geq 1$ is
a fiducial complex  number and sum over $n$. The result is \be \sum_m
B(m) e^{-\gamma m}\left(\frac{z}{z-1}\right)^{m+1} = \sum_m
\frac{F(n)}{z^n}\, , \label{z} \ee where we used the fact that  \be
\sum_n \binomial{m+n}{n} z^{-n} = \left(\frac{z}{z-1}\right)^{m+1}\, .
\ee Our next step is to expand the left--hand side of eq. (\ref{z})
into powers of $\frac{1}{z-1}$, \be \sum_{m,n\leq m+1} B(m)e^{-\gamma
m}\binomial{m+1}{n} \frac{1}{(z-1)^n} = \sum_n \frac{F(n)}{z^n}\, .
\ee Next, we multiply both sides by  $(z-1)^k,  k=0,1,..$, and perform
a   contour integral for any closed path $C$ lying outside the region
$|z| < 1$: \be \sum_m  B(m)e^{-\gamma m}\binomial{m+1}{n} \oint_C
\frac{dz}{(z-1)^{n-k}}=  \sum_{n,p\leq n} F(n)
\binomial{k}{p}(-1)^{k-p}\oint_C\frac{dz}{z^{n-p}}\, ; \ee in the
r.h.s of this equation , we have expanded the factor $(1-z)^k$ into
powers of $z$. The contour integrals of the left-- and right--hand
sides are, respectively, $2\pi i \delta_{n,k+1}$ and $2\pi i
\delta_{n,p+1}$. Thus, \be \sum_m  B(m) e^{-\gamma
m}\binomial{m+1}{k+1}= \sum_n \binomial{k}{n-1} (-1)^{k-n+1}F(n)\, .
\ee A further trick is needed to isolate $B(m)$.  Multiply both sides
of our last result by  $y^{k+1}$, where $|y| \leq 1$ is a new fiducial
quantity, and sum over $k$ ($0\leq k\leq m$) to obtain  \be \sum_n
B(n) e^{-\gamma (n+1)} \left[(1+y)^{n+1} -1\right] = \sum_n
\binomial{k}{n-1}y^k (-1)^{k-n-1}F(n) \, ; \ee then  apply the
operator $\frac{1}{(m+1)!}\left(\frac{d}{dx}\right)^{(m+1)}$ on both
sides of this equation:  \be \sum_n B(n) e^{-\gamma n+1}
\binomial{n}{m}(1+y)^{n-m}=\sum_{k,n} \binomial{k}{n-1}
\binomial{k}{m}(-1)^{k-n-1} y^{(k-m)}F(n) \, . \ee As our last step,
we take the limit $y\rightarrow -1$ and obtain \be B(m) e^{-\gamma m}
=\sum_n \binomial{k}{n-1} \binomial{k}{m} (-1)^{(m-n-1)}F(n) \, \ee
which, after substituting $n$ by $ n+1$, yields exactly eq.
(\ref{inverse}) displayed in section II.

Our next task is to evaluate

\be F(n)=-\sum_{m=0}^{\infty} e^{-\gamma m}\binomial{m+n}{n} \ln
p(m|n) \, . \ee This calculation entails a summation  of the logarithm
of a binomial, a task that we were not able to accomplish. Thus, we
resorted to the approximation
 \be \ln \binomial{m+n}{n} \approx 4 \ln 2 \frac{m n}{m+n}\, . \ee

Under this approximation, \be F(n) \approx -\sum_m
\binomial{m+n}{n}e^{-\gamma m}\left[-\gamma m +(n+1) \ln
(1-e^{-\gamma}) + 4 \ln 2 \frac{m n}{m+n}\right] \, . \ee After some
trivial manipulation, this can be rewritten as \begin{eqnarray} &
&F(n) \approx -\left[\gamma \frac{d}{d\gamma} +(n+1) \ln
(1-e^{-\gamma})\right] \sum_m \binomial{m+n}{n} e^{-\gamma m}
\nonumber\\  & & + (4 \ln 2) n \left[n e^{\gamma n}\int_\gamma^\infty
d\beta \sum_m \binomial{m+n}{n} e^{-\beta(m+n)} -
\sum_m\binomial{m+n}{n}e^{-\gamma m}\right]\, . \end{eqnarray}
Performing all the sums, the derivative and the integral, one obtains
\be F(n) \approx (n+1)\frac{\gamma e^{-\gamma}- (1-e^{-\gamma}) \ln
(1-e^{-\gamma})}{(1-e^{-\gamma})^{n+2}} -  n\frac{4 \ln 2
e^{-\gamma}}{(1-e^{-\gamma})^{n+1}} \, .
\ee

\newpage
\begin{figure}
\vspace{7cm} \special{picture FigureI} \vspace{1cm}
\caption{Approximating $f(n)=\ln\binomial{m}{n}$ by a parabola. Here,
$ m =100$.} \end{figure}

\newpage

\begin{figure} \vspace{7cm} \special{picture FigureII} \vspace{1cm}
\caption{$I_{\rm max}(x)$ for $\n = 1,10,20$ and $100$. Information is
mesured in nits.} \end{figure}

\newpage

\begin{figure} \vspace{7cm} \special{picture FigureIII} \vspace{1cm}
\caption{$I_{\rm max}(x)$ for fermions $\n = 0.5$ Information is
measured in bits.}
\end{figure}

\end{document}